# Unveiling philosophy and social aspects of nanotechnology- A short review

Muhammad Sajeer P

Center for Nanoscience and Engineering, Indian Institute of Science, Bangalore

## Abstract

Philosophy has nurtured fundamental science by asking the right questions. This scientific growth has fuelled research in various domains and introduced diverse disciplines. Nanotechnology is an interdisciplinary domain with numerous applications ranging from medical diagnostics and food technology to electronics and psychology. Exploring nanotechnology's philosophical and social perspective can better understand these domains and may open new doors for research. This review addresses philosophical and other aspects of nanotechnology, such as history, definitions, vision, language, laws, politics, and ethics. This is an attempt to equip anyone in the field of nanotechnology with philosophical and social insights. We expect this review to provide an introductory understanding of philosophy and other aspects to the nanotechnologists, which are usually excluded from their degree curriculum.

## Introduction

Nanotechnology is a comprehensive approach of studying materials and their properties on a nanometre scale by converging different scientific fields. As per the International system of units, *nano* is the one billionth of a meter. Nanotechnology coupled science and engineering and introduced a radical shift from "reporting the nature" to "rewriting the nature." This demolished the hierarchy of different scientific domains and introduced democratic interdisciplinary research. It equipped us to manipulate at the nanoscale and upgrade various technologies such as biosensors[1], cardiovascular therapeutics[2], DNA analysis[3], medically assisted reproduction[4].

A sheet of paper has a usual thickness of 100,000 nano meters. This visualisation gives insight and amusement to the smaller but exciting world of nanotechnology and its capabilities for different applications.

Unlike other technical literature, this review focus on the broader perspectives (Philosophical and social) of nanotechnology and aims to familiarise the same to nanotechnologists and vice versa. This review also attempts to emphasize that nanotechnology is a giant leap from conventional science. The philosophical outlook of nanotechnology opens new threads for research and a better understanding of the fields. Even though several philosophical questions on nanotechnology have been addressed, the majority of them remain partially answered. A few examples are the following:

- Is nanotechnology an independent branch having its own laws based on its length scale or just an extension of modern engineering Technology?
- What are the political and social implications of nanotechnology?
- Can we predict the long-term applications and impact of nanotechnology in various areas? If yes, who will be responsible for the misuse of nanotechnology?
- What is the motivating factor to do research on nanotechnology? Is it the pure interest to advance science or the business aims of investors and funding agencies? Who determines the applications or sets the milestones for nanotechnology?
- Is nanotechnology a real revolution or just hype?

# History and growth

Nanotechnology is a science based on length scales. The term "Nanotechnology" was coined by a Japanese scientist, Norio Taniguchi[5] (1974), and later popularised by Erik Drexler. The emergence of a branch in terms of length scale was remarkable in the history of science. Nanotechnology inculcates physics, chemistry, material science, and technology. Artificial intelligence and machine learning have also started contributing to its growth[6], [7].

Nanotechnology has been prevalent from earlier times. The observation of dichroism in Roman's Lycurgus cup due to silver-gold nanoparticles in the fourth century AD is an example of the same[8]. The history of modern nanotechnology is traced back to the words of Richard Feynman. He gave a lecture entitled "There is plenty of room in the bottom" at the annual meeting of the American Physical Society held at the California Institute of Technology in Pasadena on 29$^{th}$ December 1959. The lecture revealed the limitless possibilities of exploring science at the atomic scale, and his vision paved the way for modern nanotechnology [9], [10]. It is interesting to think about how he has proposed that there is plenty of space at the bottom, which accelerated the development of Nanotechnology.

The term "Nanometre" was later defined by Richard Zygmondy, who bagged a Nobel prize in chemistry in 1925. Eric Drexler and his team from the Massachusetts Institute of Technology mentioned Nanotechnology in his first paper [11] and gave credit to Feynman's vision. Drexler's perception is often considered as the base of molecular nanotechnology. "*Engines of creation: The coming era of nanotechnology*" (1986) and "*Unbounding the future: the nanotechnology revolution*" (1991), his two books have bagged attention worldwide[12], [13]. The field of nanotechnology has extended after Kroto et al. [14], and Iijima et al. [15] discovered Fullerenes and synthesized the nanotubes, respectively. The development of Scanning Electron microscopy (1937), Transmission electron microscopy (1939), Xray Photoelectron spectroscopy (1969), and Atomic Force microscopy (1982) has equipped us to visualize the nanodomain.

The excitation in nanotechnology and its potential for various applications has motivated America to put forth National Nanotechnology Initiative (NNI) after George

W Bush signed the 21st century Nanotechnology Research and Development Act[16] on 3rd December 2003. The budget of NNI was 1.2 billion US dollars in 2018. India commenced Nanotechnology research through Nanoscience and Technology Initiative with a funding of Rs. 600 million. The Nano mission, a five-year program, was also launched in 2007 with 250 million US dollars[17]. It has established national dialogues to promote research and development of nanotechnology and has laid down a National Regulatory Framework Roadmap for Nanotechnology (NRFR-Nanotech). Various countries have also initiated nanotech-related programs to utilize this field to its maximum extend. Malaysia National Nanotechnology Initiatives[18], Belgium's Nanosoc[19], and Nanologue by Germany[20] are few among them. We suggest the cited studies [21],[22] for a better understanding of the history of nanotechnology

## Terminology of Nanotechnology

The word "nanotechnology" is a combination of the terms "Nano" and "Technology." The term "Nano" is derived from the Greek word *Nanos* which means dwarf. The word "technology" comes from the two Greek words "Techne" and "logos," which means "manmade" and "language," respectively. This etymology of Nanotechnology itself shows it is not nature derived. It should be noted that nanoscience and nanotechnology are fundamentally different as the former is the science of studying materials in nanoscale, and the latter is the application of this study in different fields.

Nanotechnology is not a single entity. It is an emerging field, and several technological advances have been contributed by scientists from different disciplines worldwide. It is a compilation of different nanoscale techniques for which the word "nanotechnologies" is more precise than the word "nanotechnology." As the technologies have been contributed from different fields for different applications are different from each other, it cannot be generalized, and the term "nanotechnology" might be thus misleading. However, in this review, we will stick with the term "nanotechnology" as this is popular. A broad perspective study on terminology and nomenclature of nanotechnology has been done by Klaessig F et al. [23].

## Diversity in the definition of nanotechnology

Nanotechnology has branched out to divergent areas such as nanomedicine, nanoelectronics, nanomaterials, nanobiotechnology, and a lot more. These branches of nanotechnology are the applications of nanoscience in distinct domains ranging from electronics to medicine. The earlier philosophers have intimated that[24]; nanotechnology and its branches do not have a consensus definition. The same will be analysed further in this review in the following session. Priya Satalker has extensively studied[25] the diversity of definitions of nanomedicine and its impact on nanotechnology for clinical applications.

According to the United States nanotechnology initiative[26],

> *"Nanotechnology is science, engineering, and technology at the nanoscale, which is about 1 to 100 nanometres. Nanotechnology is the study and application of tiny things and is used across all other science fields, such as chemistry, biology, physics, materials science, and engineering."*

In their brochure, they have designated nanoscience as the "Big things from the tiny world." Royal Society and Royal Academy of Engineering have defined nanoscience and nanotechnologies in a report in 2004 as follows. [27]

> *"Nanoscience is the study of phenomena and manipulation of materials at atomic, molecular, and macromolecular scales, where properties differ significantly from those at a larger scale. Nanotechnologies are the design, characterization, production, and application of structures, devices, and systems by controlling shape and size at the nanometre scale."*

The latest definition of nanotechnology [2021] provided on the website of the Ministry of Electronics and Information Technology, Government of India, is as follows:

> *"Nanotechnology is the development and use of techniques to study physical phenomena and develop new devices and material structures in the physical size range of 1-100 nanometres (nm), where 1 nanometre is equal to one billionth of a meter. Nanotechnology impacts all areas of our lives. These include materials and manufacturing, electronics, computers, telecommunication and information technologies, medicine and health, the environment and energy storage, chemical and biological technologies, and agriculture."*

The concept and vision nanotechnology convey must be imbibed for deeper understanding in conjunction with exploring different definitions. Even though the above quoted definitions are not precisely similar, the domain of nanotechnology is clear. Some definitions introduce nanotechnology as "science deals with everything in the nanoscale" and fail to impart its ideology. Appreciative assimilation of "measurement" is also required to understand nanotechnology as it is adjoined on nano length scale." Nanotechnology is radical miniaturization, according to Eric Drexler, who popularised this term. As discussed, the definition of nanotechnology is not unanimous, and it depends on associated research and its applications.

## Ideals and visions of nanotechnology

George Khushf's work[28] on the ethics of nanotechnology suggests that there are three approaches for explaining the implications of nanotechnology. 1) A narrow-focused approach by Eric Drexler on molecular assemblers with a grand vision. 2) A broadly focused approach with zero vision proposes nanotechnology as a grab bag that includes anything and everything in nanoscale without any integrated ideals. 3) An approach that understands the factors associated with nanotechnology and thereby articulates the vision of emerging science and technology.

Even though nanotechnology is considered a pivotal concoction of science, it cannot always be accurate. Different domains have joined hands by sharing information to build novelty and have triggered the emergence of nanotechnology. It cannot be naively addressed as the overlapping of scientific branches. To better understand, scientific branches such as Chemistry and Physics exist with different ideologies and principles and cannot be overlapped perfectly. So, Nanotechnology remains as a unique discipline and not as a subset of any other areas.

Nanotechnology also has utopian and dystopian visions. It may be a problem solver which ends up as a problem creator. Its impact on living beings and the environment is still not completely understood, and researchers are actively working on it [29]–[32]. The current field of nanotechnology is the coexistence of inventions, questions, progresses, different technologies, and advantages such as cost-effective and faster functioning. It has a broader meaning, aim, and intention to advance the world exponentially and improve our lives. Nanotechnology's research direction and its vision for 2020 were discussed in Nano2 reports[33].

The discovery of a specific technology is not merely an innovation to solve a problem. Instead, it is an uncovering of the cause-effect relationship between two properties. For instance, the discovery of the telephone not only enabled the communication effortless but also revealed the cause-effect relation between electric signal and sound. Analogously, nanotechnology is not simply the properties and application of nanomaterials but a disclosure of the cause-effect relation between the size of a particle and its properties. The working space of nanotechnology lies at the brim of the classical and quantum domain. Both classical and quantum theories have been used to explain the principles of nanotechnology, and that mould it more unique and elegant.

Nanotechnology has been an absolute model of empirical knowledge. A project[34] entitled NanoforArt that uses nanotechnology techniques to restore artwork has been funded by the European Union recently and has explored the aesthetic combination of arts and science. Plant genetic materials have been manipulated to adapt to unfavourable growth conditions using nanotechnology and thereby proposed a way to demolish food scarcity. Nanotechnology has also been investigated for water treatment[35], sustainable development[36], health care[37], and a lot more. Hence, this distinctive discipline is much beyond engineering technique and has the resolution to heal mankind.

# Language of nanotechnology

The lack of adequate language to express its content and ideas is currently a loophole in nanotechnology. The nanoworld has also been explained by the same language that is exercised for macroworld. The ideas and concepts attain more clarity with the choice of terms and language used to express nanotechnology. The nano regime can be better elucidated by the term "surface to volume ratio" than the term "volume." Nanotechnology exploits a remarkably large surface-to-volume ratio of nanoparticles that gives distinct mechanical, optical, and chemical properties in contrast to the same material in bulk. To understand better, consider spherical particles in macroscale and nanoscale with diameters of 10 m and 10 nm, respectively. Their surface-to-volume ratio drastically increased from the order of $10^{-1}$ to the order of $10^8$ in the nano regime. However, the conventional method of illustration by cubes/circles/spheres to represent the macroparticles fails to convey the essence of nanoscale structures. Thus, the visualisation of concepts of nanotechnology has been demanding for the scientific community. Specific techniques or symbols to narrate the complexity of nanotechnology must be emerged to terminate the struggle for communicating the postulates of nanotechnology to the macroworld. A narrative dimension of nanotechnology can be found in Emilion Mordini's work[38].

Even though the limitless potential of nanotechnology has been employed in diverse fields such as detection of cancer[39], drug delivery[40], and semiconductor industry[41], its prospects were hard to convey by the current way of communication. Thus, proper language and narrative aspects to nanotechnology become a necessity.

## The social and ethical aspect

The ethics and social issues of nanotechnology are inevitable aspect that needs to be discussed. Nanotechnology has also been imbibed with the same distrust and fear associated with genetic engineering and nuclear energy. However, the favourable aspects of nanotechnology outweigh its disadvantages, such as the toxicity of nanoparticles. However, it has been well handled as a separate branch called nanotoxicology with proper protocols and frameworks.

Nanotechnology is not an independently evolved branch of science. Instead, it is an engineering revolution that emerged as an amalgamation of different scientific domains. Currently, there is much hype about nanotechnology as the technique of the future. It is often projected as an over-ambitious technology (or is it?). The anticipation about the broader application of nanotechnology has facilitated funding for research in its initial days. There have been even investment-related books published "*Investing in Nanotechnology*"[42] by Jack Uldrich. The Role of speculation in nanotechnology is discussed in detail by Steven Umbrello[43].

Nanotechnology has the potential for disruption as par with its potential for suitable applications. The health and environmental impact of nanotech products are extensively studied these days[44]–[46]. Trust and ethics in nanotechnology have a vital role in its constructive growth. John Weckert has discussed the trust in nanotechnology[47] in detail. Based on a survey conducted among scientists and the public, Dietram et al. has consolidated[48] the risk elements of nanotechnology. This work pointed that nanotechnology might contribute to loss of privacy, exploitation by terrorists, and cause health risks. So, it is essential to borderline the negative and positive applications of the same. A discussion on solving the issues of nanotechnology by environmental ethics can be found in the reference[49].

More than its impactful application, nanotechnology has a socio-economic dimension. These are the ethical and social implications associated with it. Scientists have coined the term "nanoethics," which corresponds to ethical, social, environmental, medical, political, economic, and legal issues related to nanotechnology. Even though nanotechnology has grown exponentially, the ethics related to it has not been explored on par with it. The social and ethical issues are also widely discussed in the literature[50]–[53]. The journals such as Nanoethics[54] focused on the ethical implications of nanotechnology. The apprehensions for much futuristic nanotechnology-enabled mind reading[55] may demolish the exponential growth of nanotechnology. A discussion on speculative nanoethics to explorative nanotechnology is done by Armin Grunwald[56]

## Politics and law

As the growth of nanotechnology is exponential, a proper toxicity study is essential to sustain human health. There are chances of nanoparticles getting trapped in the food chain, which can cause damage to the health of living beings[57]. There are political and security risks associated with nanotechnology. Technological advancements in this field can be used to develop sophisticated weapons that can harm humankind. The Institute for Soldier Nanotechnologies (ISN) by MIT utilizes nanotechnology to enhance soldiers' protection and other capabilities. Patrick Lin and his team have done a critical evaluation of nanoethics and human enhancement [58]. As nanotechnology is a comparatively new field, any government regulatory agency has limited knowledge to intervene in this[59]. The laws are inadequate to control the nanotechnology-induced environmental threats. Thus, policymakers and governments must step forward to make strong laws and regulations to ensure the safety of nanoproducts. Innovative governance of nanotechnology is vital for the development of society[60]. For example, the government recalled a German bathroom product called Magic Nano within three days as it has caused severe respiratory issues to the users.

India had issued new guidelines to ensure the safe use of nanotechnology in agriculture[61], [62]in 2019. India established its Nanotechnology and Science initiative in 2001, followed by the Nano mission in 2007. The Nanoelectronics mission was also executed by the ministry of electronics and information technology, the Government of India. Indian Nanoelectronics User Programme (INUP) has also been launched. Not only the laws and regulations about nanotechnology are in their preliminary stages, but also there are no international treaties signed for the regulation of nanoproducts. India does not have any specific law against nano toxic chemicals, contrary to the United States' Toxic Substances Control Act. However, the Environment Protection Act of 1986 facilitates ample power to the central government to draft subordinate legislation for regulating chemical substances in India. A few interesting discussions about the necessity of regulations regarding nanotechnology can be found here[63], [64].

The unique characteristics of nanotechnology enable it to be widely applicable. Nanotechnology is ubiquitous. It is utilized for food and cloth industries to medical diagnostics, and drug delivery. However, its riskiness has to be questioned. This potential of nanotechnology can be misused for terrorism. As nanotechnology can be used by governments and organizations to make miniature devices for surveillance of employees and citizens, privacy is questioned. This may also lead to political conflicts. Discussions on nanotechnology and its impact on privacy can be found in various papers[65]–[67]. A healthy politics of a country has an immense role in the advancement of nanotechnology as it requires enormous funding and approval from government bodies. It is more evident in healthcare systems[68]. Robert Sparrow has explored the ethical and political analysis of nanotechnology[69].

## Conclusion

We have discussed various aspects of the philosophy of nanotechnology. We believe this short review has provided enough evidence for the distinguishability of nanotechnology from other scientific domains and introduced a philosophical perspective of nanotechnology to scientists. Nanotechnology has broad ideals and a vision for the betterment of society. The ethical concerns should be handled so that it does not affect the rapid growth of nanotechnology. Instead of worrying too much about the far futuristic destructive implications, the bright side of nanotechnology should be appreciated. There is also an urgent need for governments to implement policies and stringent laws to prohibit its misuse. It has tremendous potential to brighten existing technologies, and hence it is a revolution. In conclusion, nanotechnology must be embraced as a significant breakthrough, provided it is democratic, and its ethical concerns are assured. We hope this review successfully provides the introductory understanding of philosophy and other aspects (Ethics, Politics, law) to the nanotechnologists, which are usually excluded from their coursework.